\documentclass[a4paper,superscriptaddress,floatfix]{revtex4}

\usepackage{epsfig}
\usepackage{amsfonts}
\usepackage{amsmath}
\usepackage{amssymb}
\usepackage{amsthm}
\usepackage{float}
\usepackage{array}
\usepackage{booktabs}
\usepackage{color}
\usepackage{graphicx}
\usepackage{hyperref}
\usepackage{multirow}
\usepackage{slashed}
\usepackage{subfigure}
\usepackage{theorem}
\usepackage{textcomp}
\usepackage[utf8]{inputenc}

\allowdisplaybreaks[1]

\numberwithin{equation}{section}

\begin{document}

\title{The effect of real and virtual photons in the di-lepton channel at the LHC}

\author{Elena Accomando}
\email[E-mail: ]{e.accomando@soton.ac.uk}
\affiliation{School of Physics \& Astronomy, University of Southampton,
        Highfield, Southampton SO17 1BJ, UK}
\affiliation{Particle Physics Department, Rutherford Appleton Laboratory, 
       Chilton, Didcot, Oxon OX11 0QX, UK}

\author{Juri Fiaschi}
\email[E-mail: ]{juri.fiaschi@soton.ac.uk}
\affiliation{School of Physics \& Astronomy, University of Southampton,
        Highfield, Southampton SO17 1BJ, UK}
\affiliation{Particle Physics Department, Rutherford Appleton Laboratory, 
       Chilton, Didcot, Oxon OX11 0QX, UK}

\author{Francesco Hautmann}
\email[E-mail: ]{hautmann@thphys.ox.ac.uk}
\affiliation{Particle Physics Department, Rutherford Appleton Laboratory, 
       Chilton, Didcot, Oxon OX11 0QX, UK}
\affiliation{Elementaire Deeltjes Fysica, Universiteit Antwerpen, B 2020 Antwerpen, Belgium}
\affiliation{Theoretical Physics Department, University of Oxford, Oxford OX1 3NP, UK}

\author{Stefano Moretti}
\email[E-mail: ]{s.moretti@soton.ac.uk}
\affiliation{School of Physics \& Astronomy, University of Southampton,
        Highfield, Southampton SO17 1BJ, UK}
\affiliation{Particle Physics Department, Rutherford Appleton Laboratory, 
       Chilton, Didcot, Oxon OX11 0QX, UK}

\author{Claire H. Shepherd-Themistocleous}
\email[E-mail: ]{claire.shepherd@stfc.ac.uk}
\affiliation{School of Physics \& Astronomy, University of Southampton,
        Highfield, Southampton SO17 1BJ, UK}
\affiliation{Particle Physics Department, Rutherford Appleton Laboratory, 
       Chilton, Didcot, Oxon OX11 0QX, UK}

 \begin{abstract}
{

We present a study of di-lepton production at the CERN Large Hadron
Collider with a particular focus on the contribution resulting from
both real and virtual photons in the initial state. We discuss the
region of phase space in which the invariant mass of the lepton pair is
of the order of several TeV, where searches for new physics phenomena
yielding a di-lepton signature are presently carried out. We study
both the yield and associated uncertainties for all possible
topologies in photon-induced di-lepton production and compare these
with what is expected in the standard Drell-Yan channel, where
quark-antiquark pairs are responsible for the production of
lepton pairs. We analyse the impact of these QED contributions on the
expected Standard Model background and on searches for new physics. In
this latter case, we use the production of an extra heavy $Z^\prime$-boson
predicted by the Sequential Standard Model (SSM) as a benchmark process.
}

\end{abstract}

\maketitle

\setcounter{footnote}{0}

\section{Introduction}

\begin{figure}[t]
\begin{center}
\includegraphics[width=0.25\textwidth]{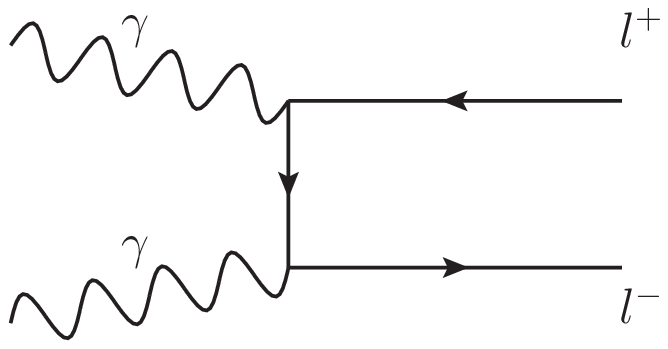}{(a)}
\includegraphics[width=0.25\textwidth]{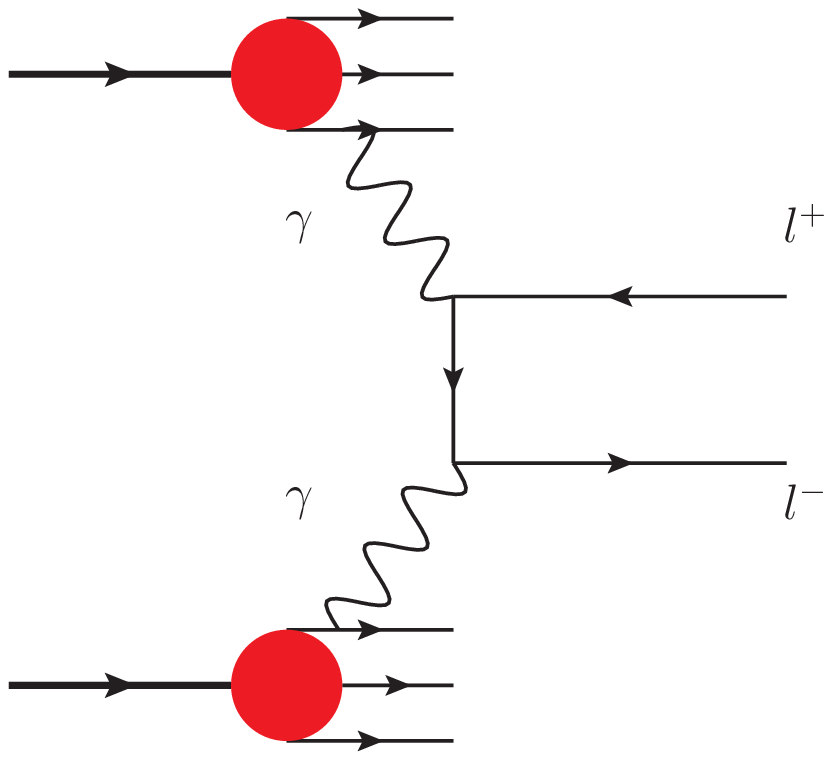}{(b)}
\includegraphics[width=0.25\textwidth]{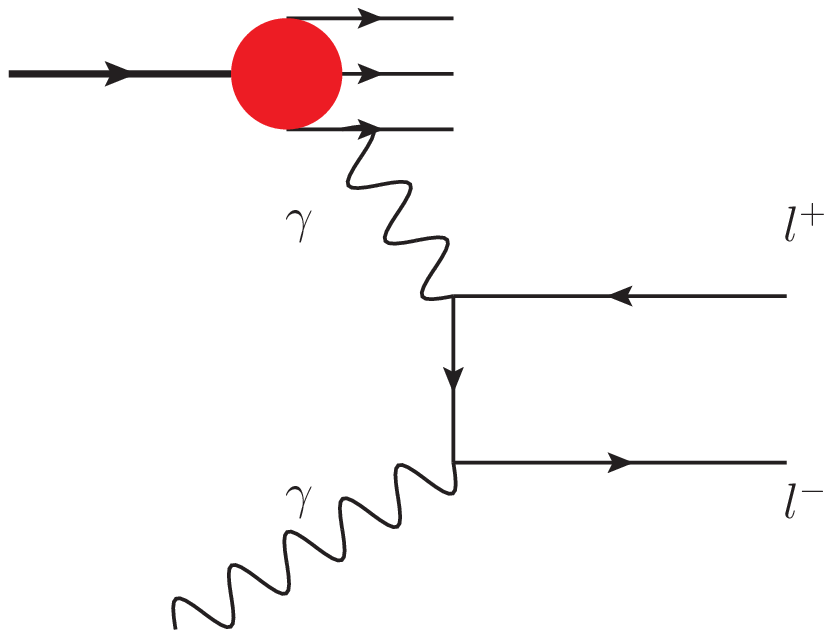}{(c)}
\caption{Photon Induced (PI) processes with (a) two real photons, (b) two virtual photons, (c) one virtual and one real photon.}
\label{fig:real_virtual}
\end{center}
\end{figure}

Photon-Initiated (PI) processes in hadron-hadron collisions can be
very important at high invariant masses
\cite{Harland-Lang:2016kog}, hence, they are relevant to 
searches for Beyond the Standard Model (BSM) processes. A prominent
example is the study of di-lepton final states, $l^+l^-$
(with $l=e,\mu$), which are used in the search
for new massive neutral gauge bosons that
are present in a variety of BSM scenarios. Currently, limits on their
masses are of the order of a few TeV \cite{Aaboud:2016cth,Khachatryan:2016zqb}. 
Alongside the standard Drell-Yan (DY) contribution,
mediated by a $s$-channel photon, $Z$ and (potentially) $Z^\prime$
bosons produced from $q\bar q$ annihilation, one has also to
account for the PI topologies shown in
Fig.~\ref{fig:real_virtual}(a). Here, photons are treated as on-shell
(i.e., massless) partons residing inside the proton and participating
in its internal dynamics governed by both Quantum Chromo-Dynamics
(QCD) and Quantum Electro-Dynamics (QED) effects (see for instance
Ref.~\cite{Martin:2014nqa}).
In this scenario, one has to define a Parton Distribution Function
(PDF) for the photon, in the same way as for (anti)quarks and gluons.
To accomplish this, different approaches are taken by different groups
and different results are obtained, both in the prediction of the
central value of the photon PDF and its error, for a given choice of
$x$ (the fraction of proton momentum carried by the photon) and $Q^2$
(the factorisation/renormalisation scale).

An analysis of such results for the hadro-production of
di-lepton pairs at the current Large Hadron Collider (LHC) was
recently presented in \cite{Accomando:2016tah, Accomando:2016ouw} (see
also Refs.~\cite{Bourilkov:2016qum,Bourilkov:2016oet}). It was pointed out that the PI
contribution can not only be sizeable but even dominant at high
invariant masses of the di-lepton pair, which is precisely where
$Z^\prime$s are searched for. The PDFs can be affected by very large
uncertainties thus generating systematics that can have an impact on
data interpretation, at high energy scales. 
\noindent
Another possible source of theoretical uncertainty comes from radiative corrections.
For the di-lepton final state in particular, these effects have been assessed in Ref.~\cite{Dittmaier:2009cr}.

The PI contribution where the two photons are both real,
i.e. massless and on-shell (hence they can be treated as proton
constituents), constitutes only one of the possible terms. 
Terms which allow either one or both photons to be off-shell 
(i.e. virtual) are also present. 
This is illustrated in Fig.~\ref{fig:real_virtual} (see plots (c) and (b), respectively). It is the
purpose of this letter to compare the phenomenology of the
three mechanisms illustrated to that of DY production of
di-lepton pairs in the high invariant mass region.

The layout of the paper is as follows: section~\ref{sec:PI} 
describes the PDF treatment of resolved photons, the EPA approach for
low-virtuality photons and the mixed process where one photon is
resolved and the other is virtual; section~\ref{sec:results}
presents our results for the Standard Model (SM) while in
section~\ref{sec:Zprime} we discuss the potential impact of PI processes
on the search for new heavy $Z^\prime$-bosons. We conclude in
section~\ref{sec:summa}.

\section{Photon induced processes}
\label{sec:PI}

In this section we discuss the three contributions in 
Fig.~\ref{fig:real_virtual} to the di-lepton
final state coming from photon induced processes. We start with the inelastic proton-proton scattering giving rise to two
resolved photons described by QED PDFs. This term is called double
dissociative (DD). Then,
we discuss the second term coming from the elastic proton-proton
scattering, where the photons are emitted with low virtuality. This
contribution is calculated in the Equivalent Photon Approximation and
is usually referred to as the EPA term. Finally we evaluate the single
dissociative (SD) contribution to the di-lepton spectrum coming from
one resolved and one low-virtuality photon.

The contribution to the di-lepton spectrum induced by two incoming real photons can be written as
\begin{equation}
\frac{d\sigma_{DD}}{dM_{\ell\ell}} = \iint dx_1 dx_2 \frac{1}{32\pi M_{\ell\ell}} \left|\mathcal{M}(\gamma\gamma\rightarrow l^+l^-)\right|^2 f_\gamma(x_1, Q)f_\gamma(x_2, Q)
 \label{eq:double_diss}
\end{equation}
where the function $f_\gamma(x, Q)$ is the photon PDF, the variables
$x_{1,2}$ are the fraction of proton energies taken away by the two
photons and $Q$ is the factorisation scale. The observable
$M_{\ell\ell}$ is the di-lepton invariant mass. The PDF error
is estimated in different ways by different collaborations. NNPDF3.0
provides uncertainty estimates following the ``replicas" method
\cite{Ball:2011gg}, while CT14QED follows the Hessian procedure
\cite{Stump:2003yu}, giving a table of 31 PDFs each one associated
with a fixed amount of the proton energy fraction carried by the
photon. We estimate the error on the Double-Dissociative contribution
to the di-lepton production, following the different approaches
adopted by the various PDF collaborations (see
Ref.~\cite{Accomando:2016tah, Accomando:2016ouw} for details). 
The more recent LUXqed set~\cite{Manohar:2016nzj} has been delivered in the LHAPDFv6 format~\cite{Buckley:2014ana}, and is accompanied by a set of 100 symmetric Hessian eigenvalues to estimate the PDF uncertainties, following the procedure explained in Ref.~\cite{Butterworth:2015oua}.
The MRST2004QED collaboration does not provide a recipe for estimating an error. Therefore 
we only show the results for the central value in this case.

The EPA is used in the literature to evaluate the contribution to the
di-lepton final state that comes from two initial low-virtuality
photons. In the EPA,
the photon flux of the proton is estimated by semi-empirical formulae
built on the dipole approximation, whose parameters are fitted to the
deep-inelastic electron-proton scattering
data~\cite{Budnev:1974de}. The photon induced differential cross
section is obtained by multiplying the photon luminosity by the matrix
element of the hard photon-photon subprocess and integrating over the
phase space (for details see for example
Ref.~\cite{Piotrzkowski:2000rx}):

\begin{equation}
\frac{d\sigma_{EPA}}{dM_{\ell\ell}} = \frac{dL_{\gamma\gamma}}{dM_{\ell\ell}} \sigma_{\gamma\gamma} = \int_{Q^2_{1,min}}^{Q^2_{1,max}} dQ^2_1 \int_{Q^2_{2,min}}^{Q^2_{2,max}}dQ^2_2 \iint dx_1dx_2 \frac{\left|\mathcal{M}(\gamma\gamma\rightarrow l^+l^-)\right|^2}{32\pi M_{ll}}N(x_1,Q^2_1)N(x_2,Q^2_2)
\end{equation}
Here, $Q^2_{min}$ is determined by the kinematics whilst $Q^2_{max}$
is arbitrary. This is the main source of error
\cite{Kessler:1994ic}. In order to estimate the uncertainty on the EPA
predictions, we have spanned the interval
$0.5~{GeV}^2 < Q^2_{max} < 8~{GeV}^2$. Our implementation has been found to agree
with that in Ref.~\cite{Schul:2011xwa}. The bulk of the effect comes
from photons with low virtuality, while photons with a
virtuality $>2~{GeV}^2$ do not give an appreciable contribution.

\begin{figure}[t]
\begin{center}
\includegraphics[width=0.47\textwidth]{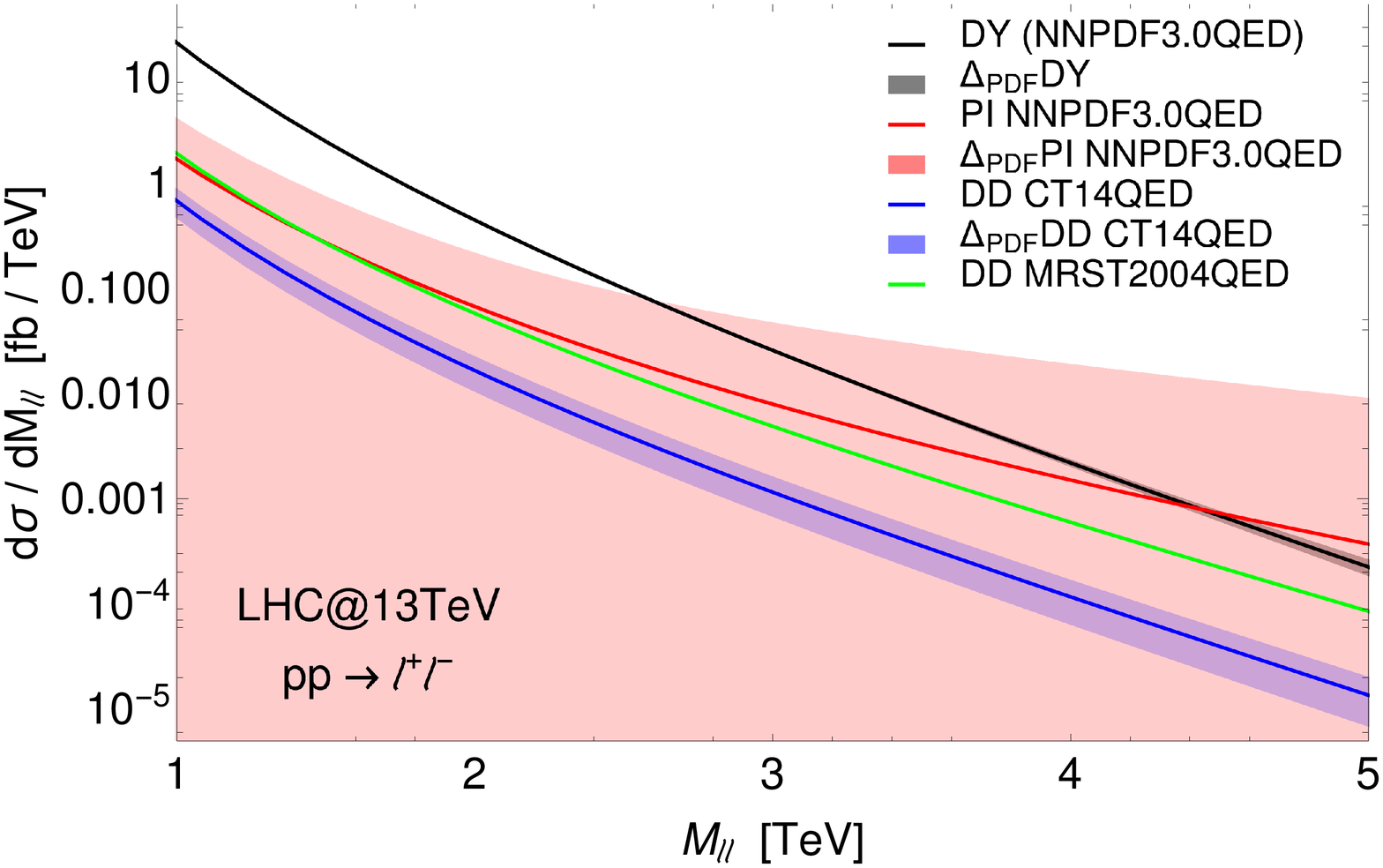}{(a)}
\includegraphics[width=0.47\textwidth]{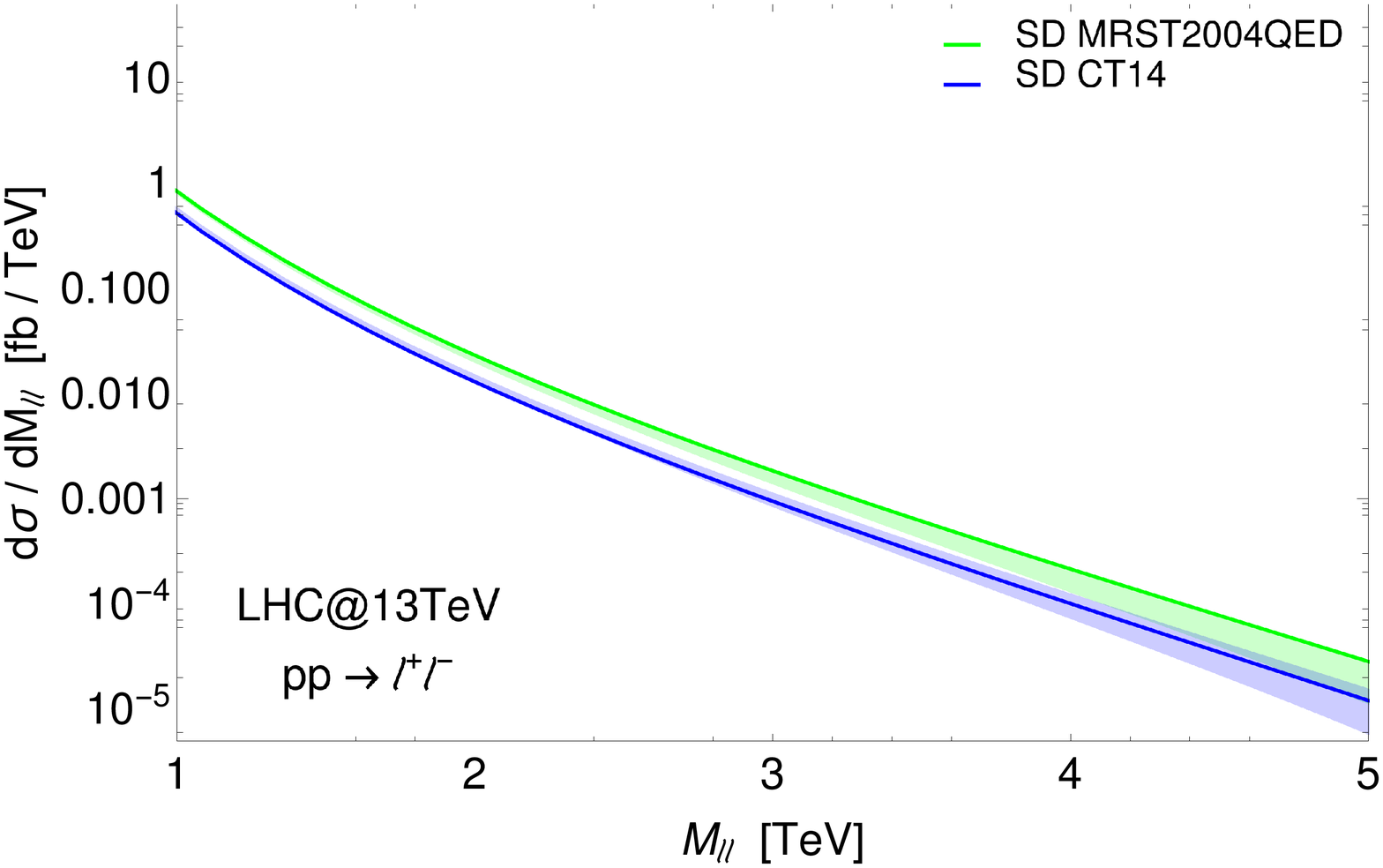}{(b)}
\includegraphics[width=0.47\textwidth]{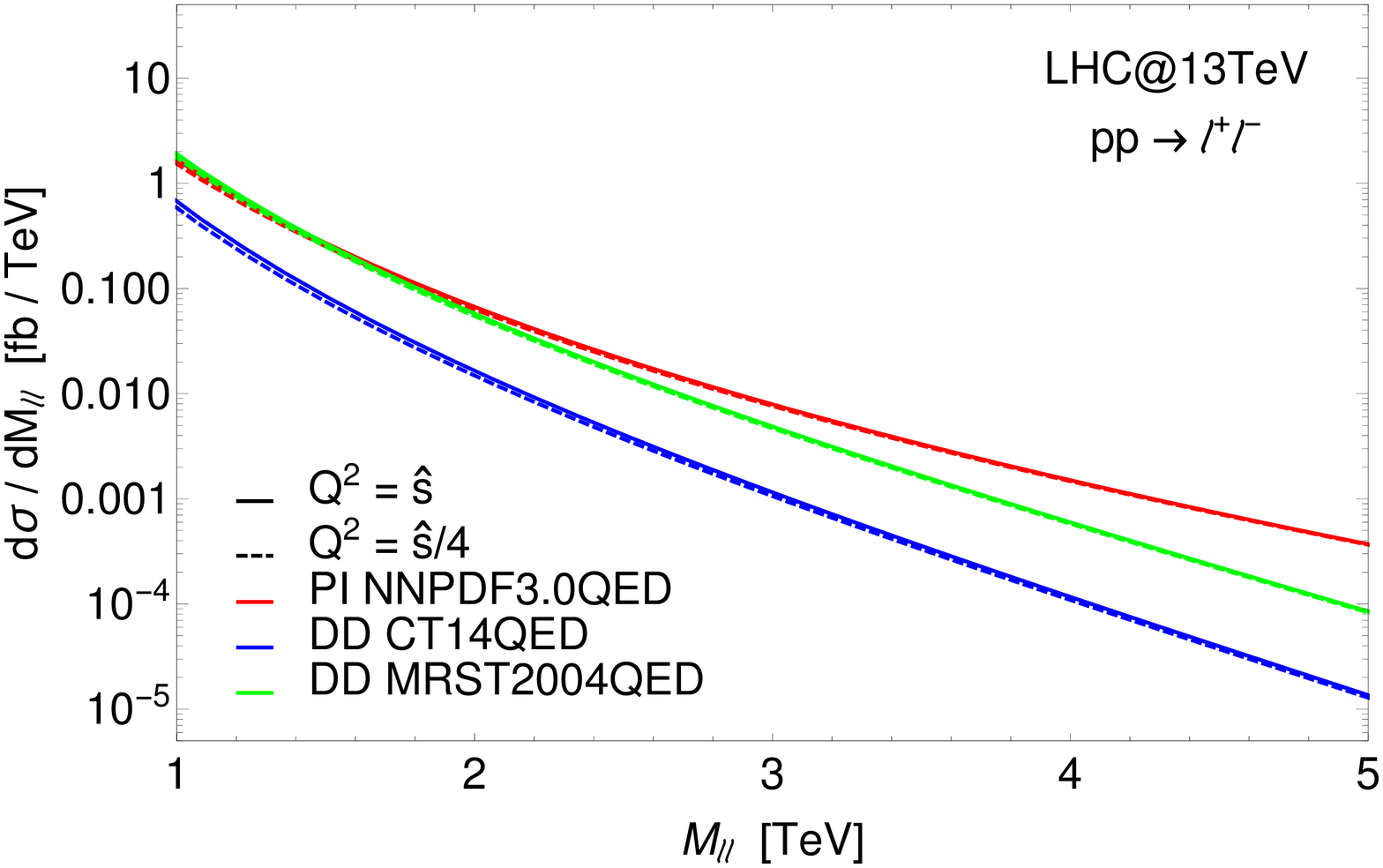}{(c)}
\includegraphics[width=0.47\textwidth]{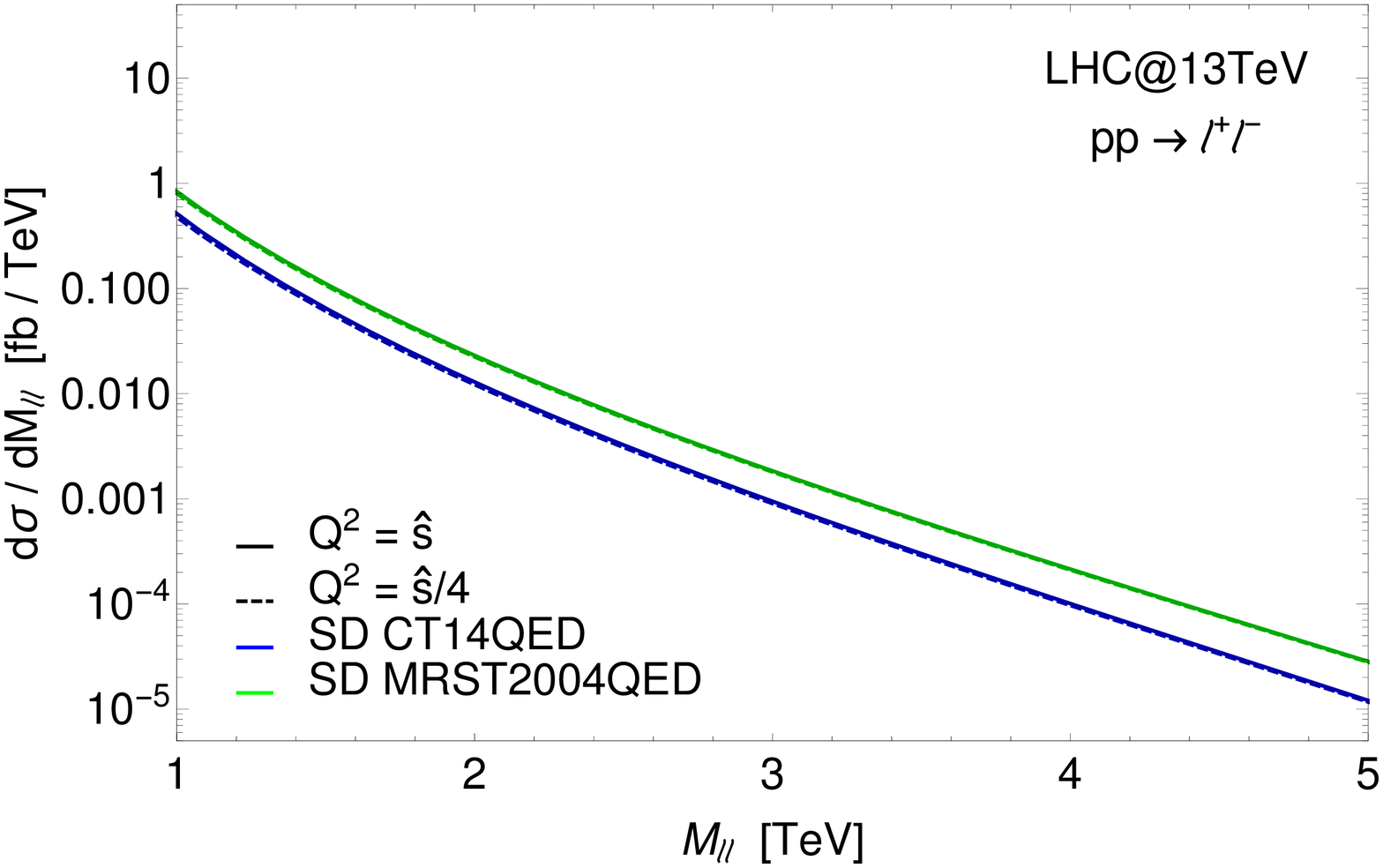}{(d)}
\caption{(a) Central value (solid line) and error (shaded area) of the Double-Dissociative (DD) contribution (or the inclusive PI result in the NNPDF3.0QED case) to the di-lepton spectrum at the LHC@13TeV. The red curve and shaded area is the NNPDF3.0QED prediction, the blue curve and shaded area is the CT14QED one, the green line comes from MRST2004QED. The black curve shows the DY contribution taken as a reference. Standard acceptance cuts are applied: $|\eta_l| < 2.5$ and $p_T^l > 20$ GeV. 
(b) Same as (a) for the Single-Dissociative (SD) contribution.
(c) Predictions for the central value of the DD term obtained for two choices of factorization scale.
(d) Same as (c) for the SD term.} 
\label{fig:DD-SD}
\end{center}
\end{figure}

Real photons extracted from the QED PDFs of one proton can interact
with the low-virtuality photons of the other proton, producing a
real-virtual photon hybrid interaction that is referred to as the
Single Dissociative (SD) contribution to the di-lepton final state.
In order to estimate this term we integrate over one
EPA flux and one photon PDF:
\begin{equation}
\frac{d\sigma_{SD}}{dM_{\ell\ell}} = \int_{Q^2_{1,min}}^{Q^2_{1,max}} dQ^2_1\iint dx_1dx_2 \frac{\left|\mathcal{M}(\gamma\gamma\rightarrow l^+l^-)\right|^2}{32\pi M_{ll}} N(x_1,Q^2_1)f_\gamma(x_2, Q)+(x_1\leftrightarrow x_2)
 \label{eq:single_diss}
\end{equation}
where $Q$ is the factorisation scale appearing in the resolved photon
PDFs. The last sum accounts for the multiplicity of the process
(virtual-real + real-virtual). The predictions for this term are of
course PDF dependent and we will show results using different 
PDF sets. In order to estimate the systematic error on the SD term, we
consider both the variation of $Q^2_{1,max}$ on the contribution from
the virtual photon, and the PDF error on the resolved photon. 

\section{QED effects on the SM di-lepton spectrum}
\label{sec:results}

The double dissociative contribution to the di-lepton spectrum at high
energy scales has been recently analysed by different authors
\cite{Bourilkov:2016qum,Accomando:2016tah,Accomando:2016ouw} for LHC
centre of mass energies of 8TeV and 13 TeV. The effect of this
photon-photon interaction has also been studied for top-antitop
pair production \cite{Hollik:2007sw,Pagani:2016caq,Tsinikos:2016xwn}
at the LHC and strong similarities with the lepton channel were 
found. In the literature, the main focus is on the DD term and its
uncertainty. There exist many photon PDF sets in the
literature: MRST2004QED \cite{Martin:2004dh}, CT14QED
\cite{Schmidt:2015zda}, NNPDF2.3QED \cite{Ball:2013hta,Ball:2014uwa}, APFEL-NN2.3QED \cite{Bertone:2013vaa}, NNPDF3.0QED \cite{Bertone:2016ume}, LUXqed
\cite{Manohar:2016nzj}, xFitter\_epHMDY \cite{Giuli:2017oii}. 

\begin{figure}[t]
\begin{center}
\includegraphics[width=0.47\textwidth]{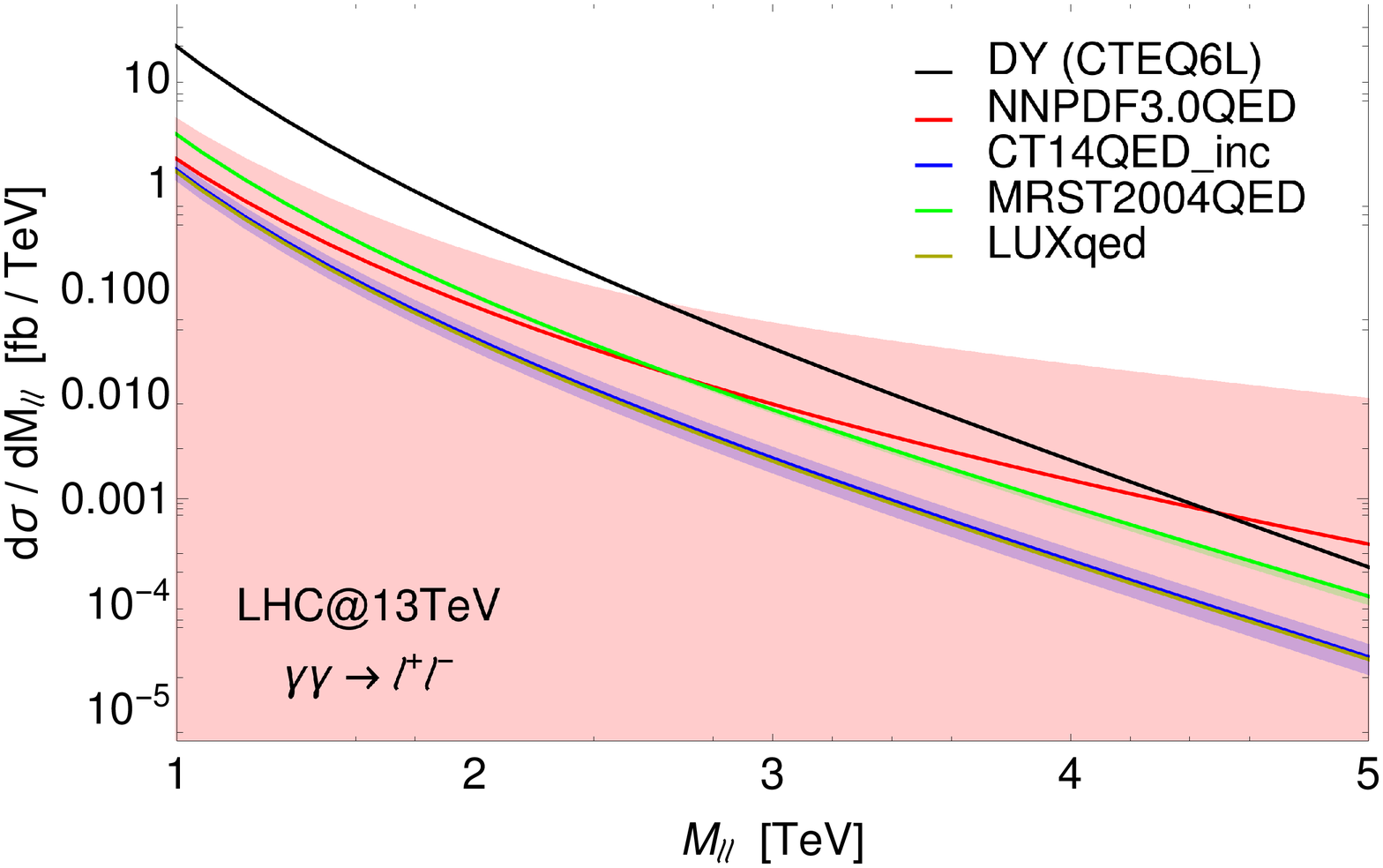}{(a)}
\includegraphics[width=0.47\textwidth]{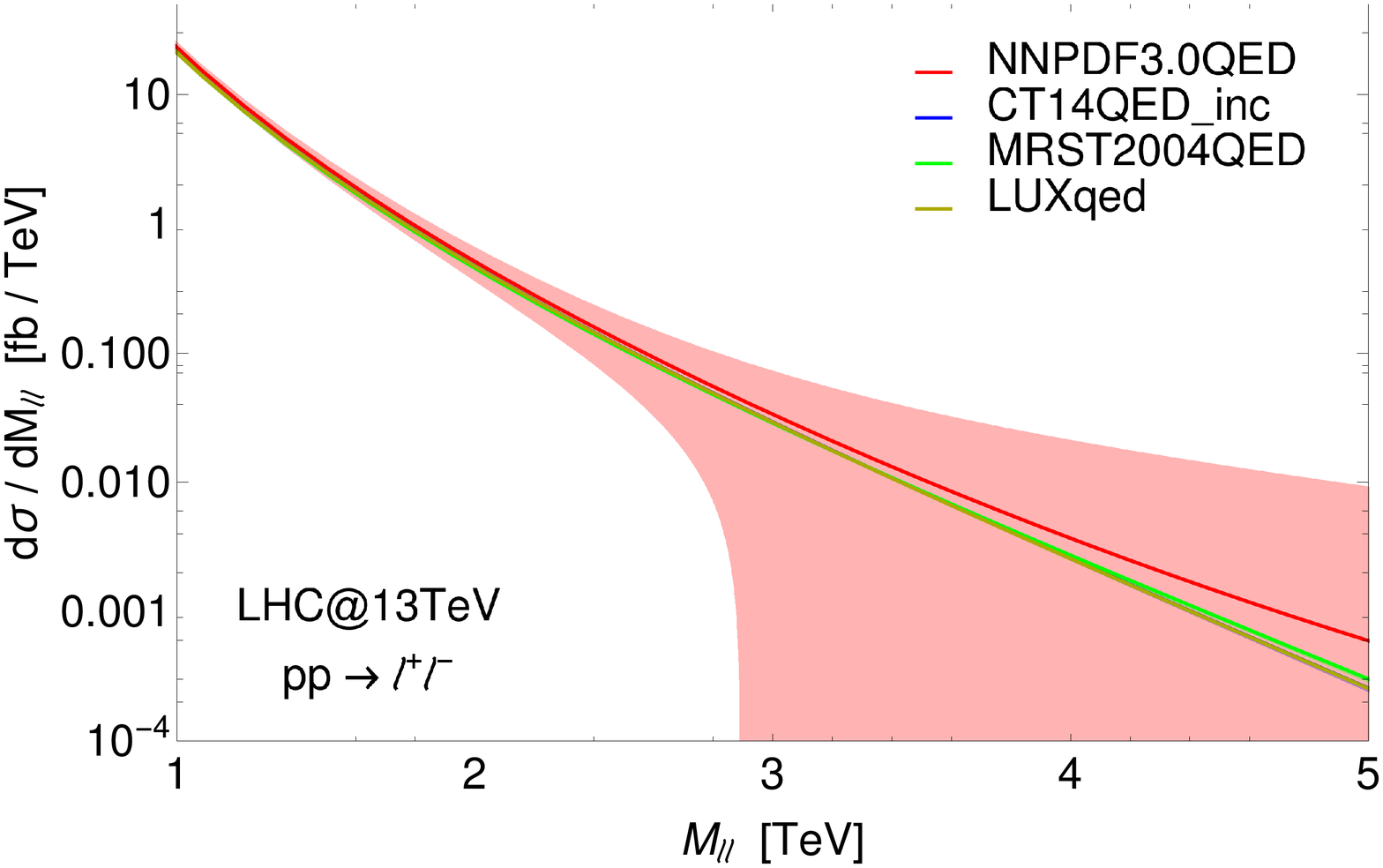}{(b)}
\includegraphics[width=0.47\textwidth]{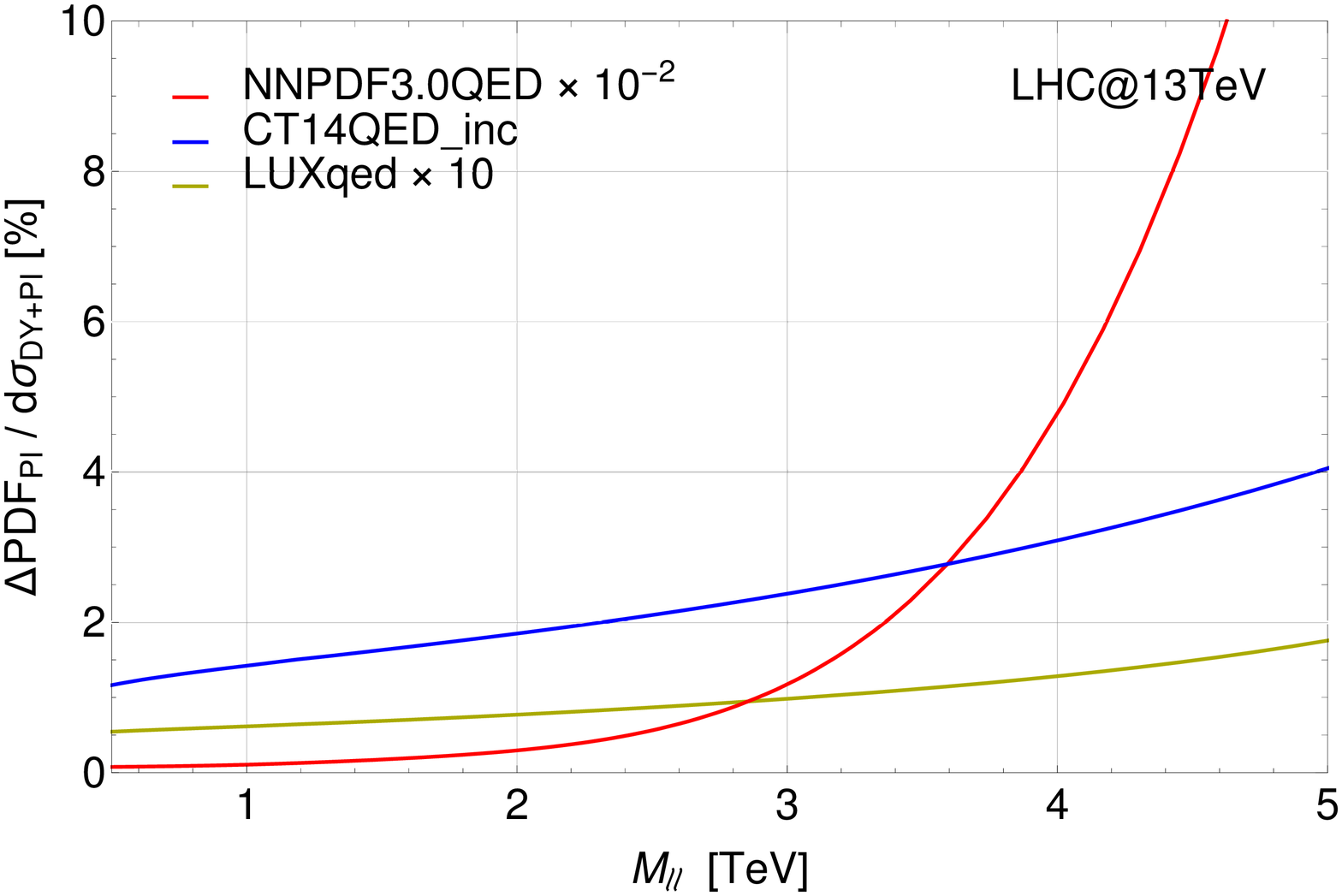}{(c)}
\includegraphics[width=0.47\textwidth]{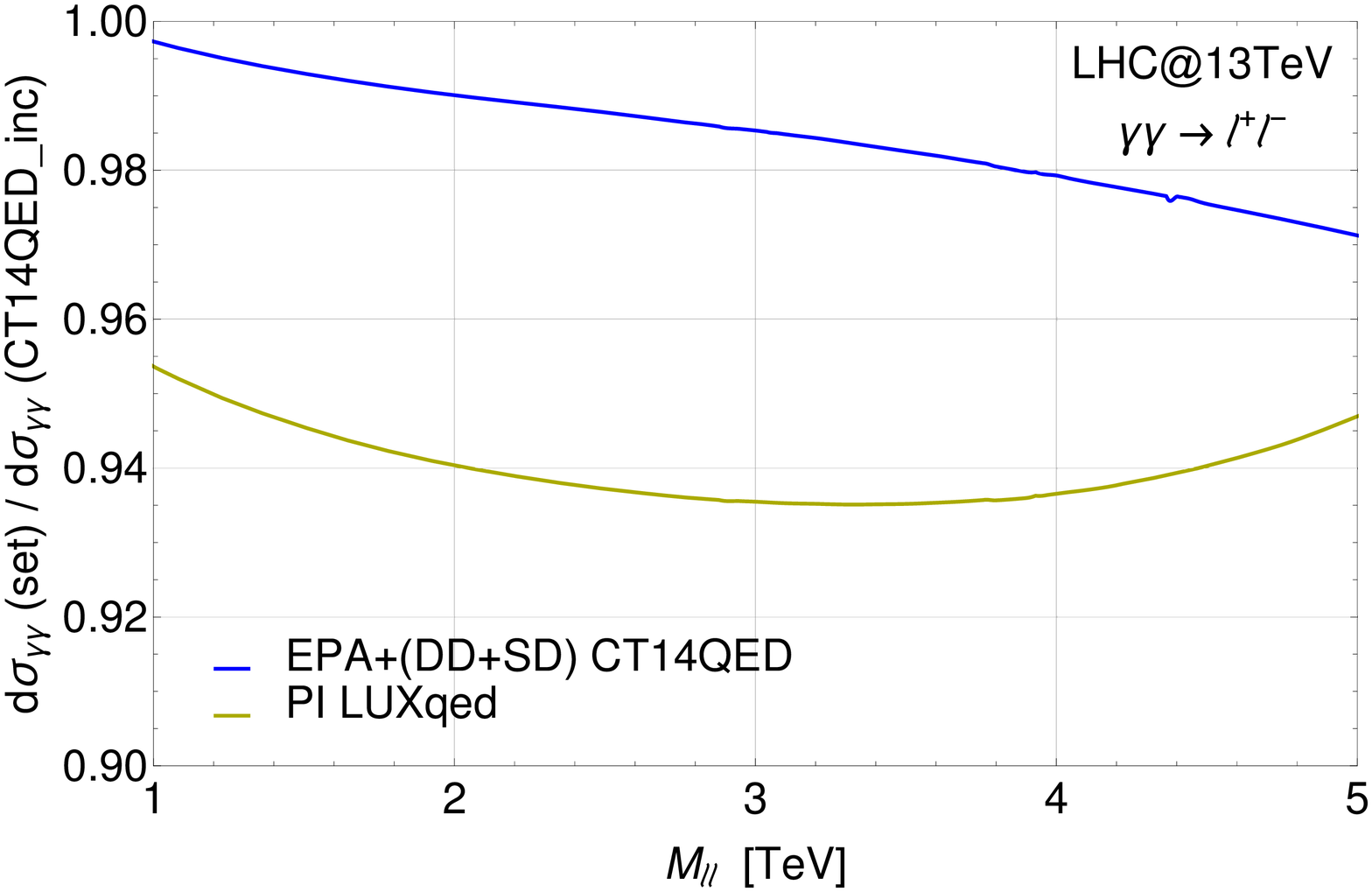}{(d)}
\caption{(a) Inclusive results for the various PDF sets and the sum of the EPA, DD and SD terms for the MRST2004QED set, in comparison with the pure DY term (here the LUXqed curve overlaps the CT14QED\_inc one).
(b) We include the DY contribution in the results obtaining the complete prediction for the di-lepton invariant mass distribution from the various PDF sets. The shaded areas represent the photon PDF systematic uncertainty. 
(c) Relative size of the photon PDF error over the complete di-lepton spectrum (PI+DY). The DY has been evaluated using CTEQ6L set for the LUXqed and CT14QED\_inc curves, while for the other curve we have used the NNPDF3.0QED set only. Note that the curves have been rescaled accordingly to the legend in order to fit in the plot.
(d) Ratio between the PI cross sections obtained with the LUXqed  and the CT14QED\_inc sets, and between the sum of the EPA, DD and SD terms obtained with the CT14QED set and the inclusive result of CT14QED\_inc.
}
\label{fig:dilepton_tot}
\end{center}
\end{figure}

Some of the PDF sets, such as CT14QED\_inc and LUXqed, include the elastic component from 
quasi-real photons, while in others, such as CT14QED and MRST2004QED, this is subtracted off. 
Using the approach we have set out in Section~\ref{sec:PI}, we are in a position to check the consistency between this subtraction procedure 
and our calculations. This will be shown in Fig.~\ref{fig:dilepton_tot}(d) where we will compare the sum of the three terms obtained with the CT14 
set, with the inclusive result obtained with CT14QED\_inc, finding a good agreement.
In Fig.~\ref{fig:dilepton_tot} we will also examine the case of LUXqed, and find that it gives 
results close to those of CT14QED\_inc when evaluating the di-lepton spectrum at high scales.  
The sets NNPDF3.0QED and xFitter\_epHMDY do not attempt 
to separately model elastic and inelastic photon components. They perform fits to 
data inclusively, and this results into much larger uncertainties than CT14 and LUXqed.   
The NNPDF3.0QED set is the last release of the NNPDF collaboration including photon PDFs. In order to analyse the leptonic 
pair production at the LHC RunII, in Fig.~\ref{fig:DD-SD}(a) we therefore choose to show the predictions for the photon induced contribution 
to the di-lepton spectrum coming from the three representative PDF sets MRST2004QED, CT14QED, NNPDF3.0QED.   

For the first two PDF sets we show separate results for the DD and the SD term, while for the latter we show the inclusive result, 
and we compare these contribution to the SM
Drell-Yan (DY) spectrum. For each prediction, we include the
associated error bands, estimated as discussed above. We apply
standard acceptance cuts on rapidity and transverse momentum of the
two final state leptons: $|\eta_l| < 2.5$ and $p_T^l > 20$ GeV (see Ref.~\cite{Khachatryan:2014fba}). We find that the three PDF sets agree quite well for the DY channel prediction plus uncertainty. The three curves in fact overlap within the NNPDF3.0 error band, which in the DY case is
quite narrow. The effect of the two initial real photons has instead different outcomes. The prediction from CT14QED is by far the smallest one, followed by MRST2004QED. The PI central value given by NNPDF3.0QED is overwhelming the MRST2004QED and CT14QED predictions. The agreement between the various PDF collaborations is not optimal, if we look at the central
values. The discrepancy between the fit driven CT14QED and the
modelled MRST2004QED is also significantly bigger than the CT14QED
error band. However, the NNPDF3.0QED error on the spectrum is so large at
high scales that one can conclude that all PDF sets are in good
agreement with each other, within that assumed accuracy. The
consequence of this assumption is that we are in the presence of a very large 
theoretical systematics at high scales. For $M_{ll} >$ 3 TeV, the
theoretical uncertainty on the leptonic DY channel, using the NNPDF3.0QED set, is of the order $O(100\%)$ and more.

\begin{figure}[t]
\begin{center}
\includegraphics[width=0.47\textwidth]{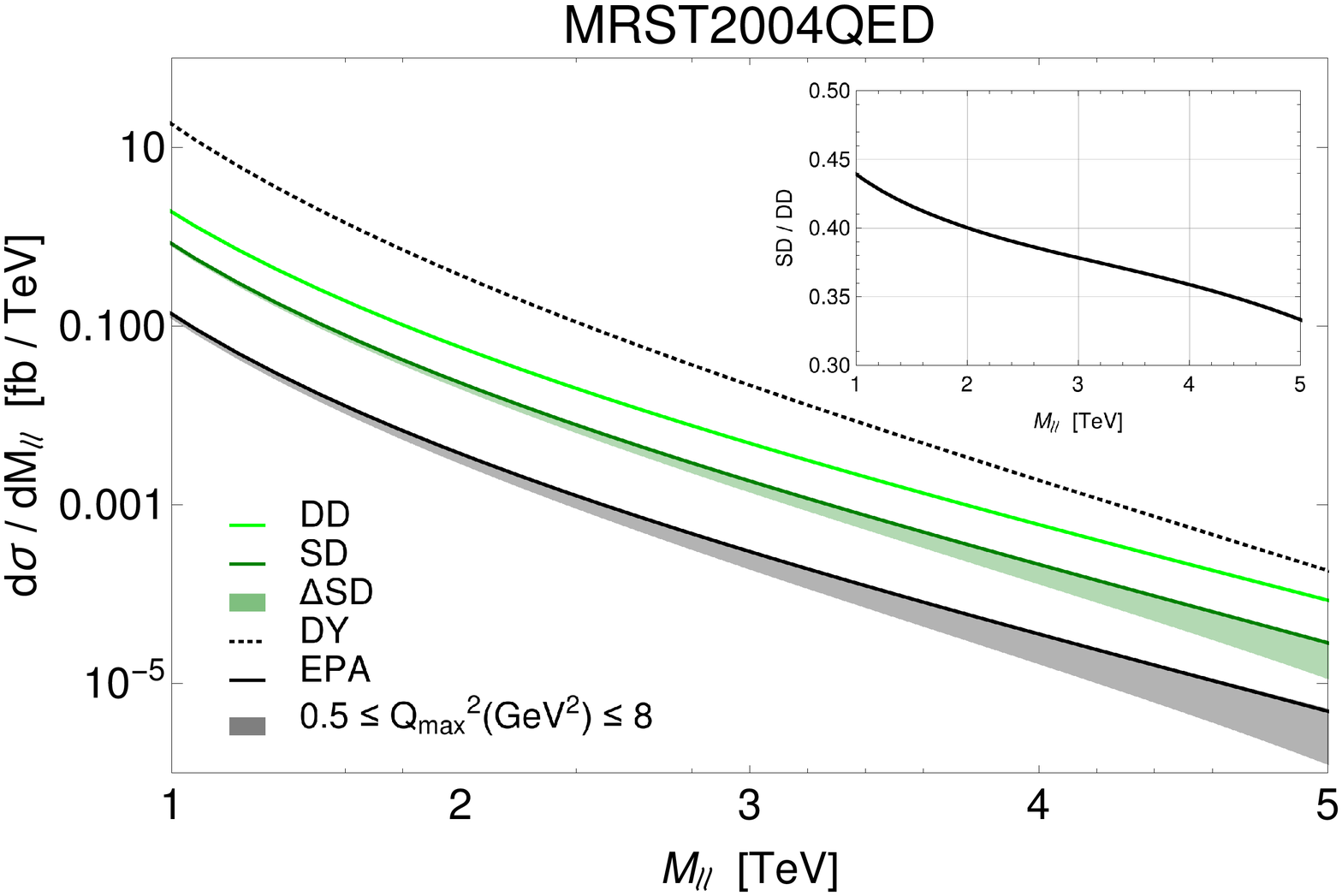}{(a)}
\includegraphics[width=0.47\textwidth]{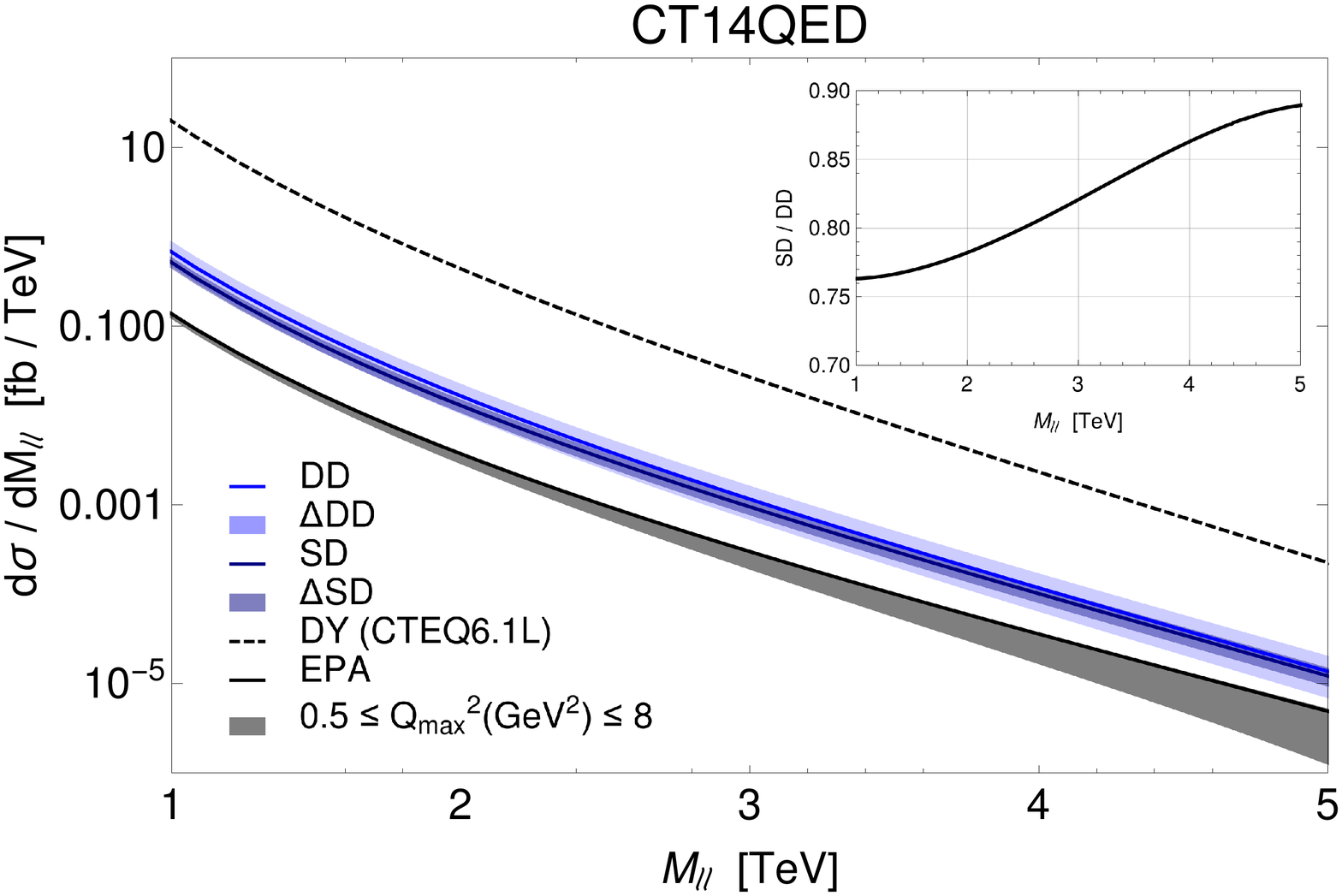}{(b)}
\caption{(a) Individual photon-induced contributions to the di-lepton spectrum at the LHC@13TeV, computed with the MRST2004QED PDF set. From bottom to top, we show the contribution from two virtual photons in EPA (black solid line and shaded area), from one real and one virtual photon (SD) (darker coloured line and shaded area), and from two real photons described by QED PDFs (DD) (light coloured line and shaded area). The top black dashed line is the reference DY spectrum. The top-right inset plot shows the ratio between the SD and DD contributions as a function of the di-lepton invariant mass. (b) same as (a) with the CT14QED PDF set. The shaded areas represent the error coming from photon PDFs and/or from the $Q^2_{max}$ choice in EPA.}
\label{fig:PI_EPA_single}
\end{center}
\end{figure}

\noindent

In this paper we examine the SD and EPA contributions to the di-lepton
spectrum at the LHC to investigate their possible relevance at high
invariant masses. Interestingly, the SD term is found to be sizeable and not
extremely sensitive to the chosen PDF set. 
Also the same hierarchy as the DD term when varying PDFs is displayed, but to a much
lesser degree, as visible in Fig.~\ref{fig:DD-SD}(b) where we
compare the predictions for the SD term given by the two selected
PDF sets (the colour code is the same as in Fig.~\ref{fig:DD-SD}(a).).
For completeness we also show the effect of different choices of the factorization scale in Fig.~\ref{fig:DD-SD}(c) and Fig.~\ref{fig:DD-SD}(d) respectively for the DD (or the fully inclusive result for the NNPDF3.0QED choice) and the SD terms. 
The effect of scale variation has also been explored in Ref.~\cite{Accomando:2016tah}, where it is visible that for the choice $Q^2 = p_T^2$ as well we do not observe significant deviations.
\noindent
The EPA term is displayed in Fig.~\ref{fig:PI_EPA_single} and found to
be negligible. This figure shows the three photon-induced 
contributions to the di-lepton spectrum for a centre of mass energy of 
13 TeV at the LHC, and we compare them to the DY channel, for the two non-inclusive PDF sets separately.
In Fig.~\ref{fig:PI_EPA_single}(a), we show the results
obtained by using the MRST2004 set. From bottom to top, we display the
contributions to the spectrum that come from two initial low-virtuality
photons treated in EPA (black curve and shaded area), the
SD interaction between a real and a low-virtuality photon (deep green
curve and shaded area) and the DD interaction between two
initial resolved photons. The solid lines give the central value while
the shaded areas represent the PDF uncertainty, when available. The
black dashed line on top of the plot is the pure DY channel at the QCD
and EW lowest order, generated by the quark-antiquark interaction. The
EPA term is subdominant compared to the SD and DD contributions. As to
these latter ones, the top-right inset plot of
Fig.~\ref{fig:PI_EPA_single}(a) shows the ratio between the single and
double dissociative central values. The ratio decreases with
increasing di-lepton invariant mass. Beyond the 2 TeV region, of
interest for BSM searches, the SD contribution is 30 - 40\% of the DD
one. In this case, the error on the SD term comes only from the
treatment of the virtual photon in EPA as the MRST2004 set does not
provide any PDF uncertainty.

\noindent
In Fig.~\ref{fig:PI_EPA_single}(b), we show the same photon induced
contributions to the di-lepton spectrum evaluated using the CT14QED
set of PDFs. The EPA contribution (black curve and shaded area) is
obviously unchanged, as it doesn't involve PDFs. Following the color
code of Fig.~\ref{fig:DD-SD}, we display the SD term (deep blue curve
and shaded area) and the DD term (light blue curve and shaded
area). The reference DY spectrum is given by the top curve (black
dashed line), computed with CTEQ6L. The SD term shows the same
behaviour as the DD one, when comparing MRST2004QED with CT14QED. The
CT14QED result is in fact slightly smaller than the MRST one over the
entire mass range. However, in contrast to the MRST2004 result given
in Fig.~\ref{fig:PI_EPA_single}(a), the magnitude of the SD
contribution is now of the same order as the DD one. The two
contributions overlap within the DD error band. This is explicitly
shown in the top-right inset plot. There, one can see that the ratio
SD/DD is of the order of 80 - 90\% for $M_{ll} >$ 2 TeV. Both the
magnitude of the ratio of the SD to DD terms and the behaviour of this
ratio as a function of the energy scale differ between CT14QED and
MRST2004QED. For the first PDF this ratio rises and for the second it
falls. These two features are actually due almost entirely to the
difference in the DD term that the two collaborations have whereas the
two SD predictions almost agree to within the error shown. 

\noindent

In summary, the EPA
contribution can safely be neglected since it is generally two orders of
magnitude smaller than the DY differential cross section. The SD terms
given by the different collaborations are in much better agreement than
the DD ones. Moreover, the CT14QED result agrees within errors with
MRST2004QED, in contrast to the DD case. 
Even if the SD central values are very similar, the relative impact of the SD term
on the total photon induced process depends strongly on the PDF set
that is used, as the DD central values predicted by the different
collaborations are quite different. As before, the main conclusion is
that the SD term generates an additional error to the already large systematic
uncertainty in the high invariant mass region of the di-lepton
spectrum.

\noindent

In order to visualise the overall effect due to initial photon
interaction, in Fig.~\ref{fig:dilepton_tot}(a) we plot the complete
photon induced contribution to the di-lepton spectrum obtained by
summing up the EPA, SD and DD terms for the MRST2004QED set and we compare it to the default
DY differential cross section and to the inclusive PI predictions obtained with the NNPDF3.0QED, CT14QED\_inc and LUXqed sets . 
The complete prediction for the invariant mass distribution of the di-lepton pairs including the
overall PDF uncertainties for each PDF set is visible in
Fig.~\ref{fig:dilepton_tot}(b). 
There is a substantial overlap between the CT14QED\_inc and the LUXqed results such that they cannot be disentangled in this plot scale.
The prediction of the theoretical errors is quite different in the various PDF set, and some of them are not visible in the previously discussed plot scales.
In Fig.~\ref{fig:dilepton_tot}(c), using the rescaling factors visible in the legend, we show the relative size of the photon PDF error with 
respect to the complete result, sum of the PI and the DY contributions. 
The most accurate prediction is given by the LUXqed set, where the photon PDF relative error is always below 0.2\% over the entire spectrum.
The most conservative scenario is given by the NNPDF3.0QED set, where the photon systematics grow above 10 times the central value 
in the high invariant mass region.
The CT14QED\_inc has an error lying between 1\% and 4\% along the spectrum.
If we consider the predictions of the CT14QED\_inc or LUXqed, the PI contribution can be safely neglected in  BSM searches based  on the di-lepton spectrum.
In the NNPDF3.0QED case instead, the PI contribution and its uncertainty are statistically
relevant already at the ongoing LHC RunII.
With the present collected
luminosity $\mathcal{L}\simeq 13 fb^{-1}$, one would expect zero SM
background events beyond $M_{ll}$ = 3 TeV. If taking into account the
NNPDF3.0QED  uncertainty, one could  possibly have up to 3 events
at that mass scale. 
This result could have a profound impact on BSM searches for wide resonances and in particular non-resonant scenarios, like contact interactions, where the SM background is estimated from theory as we discuss in the next section.

As a cross check to exclude possible double counting intrinsic in the QED PDFs definition, 
we have compared the PI contribution obtained through eq. \ref{eq:double_diss} using the inclusive CT14QED\_inc set, 
with the sum of the three separate terms EPA, SD and DD obtained using the CT14QED set. 
The result is shown in Fig. \ref{fig:dilepton_tot}(d) where we plot the ratio between these two quantities (blue line).
Here we can appreciate that using different PDF definitions we are  able to separate inclusive and exclusive contributions, keeping the double counting at the level of percent or below. 
In the same plot we show the comparison between the two close central values for the PI obtained with the LUXqed and the CT14QED\_inc sets, their differences being below 7\% in all the spectrum.

In the next section we examine implications of the above results on $Z^\prime$-boson searches 
at the LHC. As noted earlier, the LUXqed set is the set which gives the smallest photon PDF uncertainties, 
while NNPDF3.0QED gives the largest. Taking the LUXqed uncertainty bands, the central values both for sets which employ a 
very different approach from LUXqed (such as NNPDF3.0QED and xFitter\_epHMDY) and for sets which are more similar in spirit to LUXqed 
(such as MRST2004QED and CT14QED) are outside the uncertainty bands. 

\section{QED effects on $Z^\prime$-boson searches}
\label{sec:Zprime}

In this section, we will explore the impact of the photon induced
processes on searches for $Z^\prime$ resonances in
the di-lepton channel for an LHC centre of mass energy of 13 TeV. For
a review of $Z^\prime$-boson physics see
Refs.~\cite{Langacker:2008yv,Erler:2009jh,Nath:2010zj,Accomando:2010fz}
and references therein. 

In this paper we adopt the recently published NNPDF3.0 for computing both the
$Z^\prime$-boson signal and the SM background taking into account the photon
induced contributions. The earlier paper on this topic in 
Ref.~\cite{Accomando:2016tah} used the NNPDF2.3 set.
While there is no observable change in the $Z^\prime$-boson 
signal and the SM DY background, both induced by quark-antiquark
interactions, the QED terms have been improved. The NNPDF3.0QED set has
significantly reduced the global photon induced central value 
by a factor of 1.7 - 1.8 for $M_{ll} > $ 3 TeV. Moreover, it has
slightly decreased the error band by roughly 20\% in the same invariant
mass region, while still preserving the agreement between all PDF 
collaborations.
\noindent
We use inclusive sets results to examine the consequences of photon-induced 
contributions to the di-lepton spectrum on $Z^\prime$-boson
searches at the LHC.
As already shown in
Ref.~\cite{Accomando:2016tah}, bump searches for narrow resonances are
not significantly affected by the systematic uncertainties due to
initial photon interactions. This is because these searches are not extremely 
sensitive to the background shape. It is in the event of searches for wide
resonances or for non-resonant new physics, like contact interactions (CI),
that systematic uncertainties can play a role. To illustrate 
the possible impact of such an uncertainty, we assume that the expected SM background is given by the theoretical prediction. 
For the CI search, one does have to assume the theory prediction, indeed, whilst for a wide resonance it may not be so constrained. 
In this paper, we take as benchmark theory the Sequential Standard Model
($SSM$) discussed in Ref.~\cite{Altarelli:1989ff} with a reasonably rather wide width. 

\noindent

\begin{figure}[t]
\begin{center}
\includegraphics[width=0.45\textwidth]{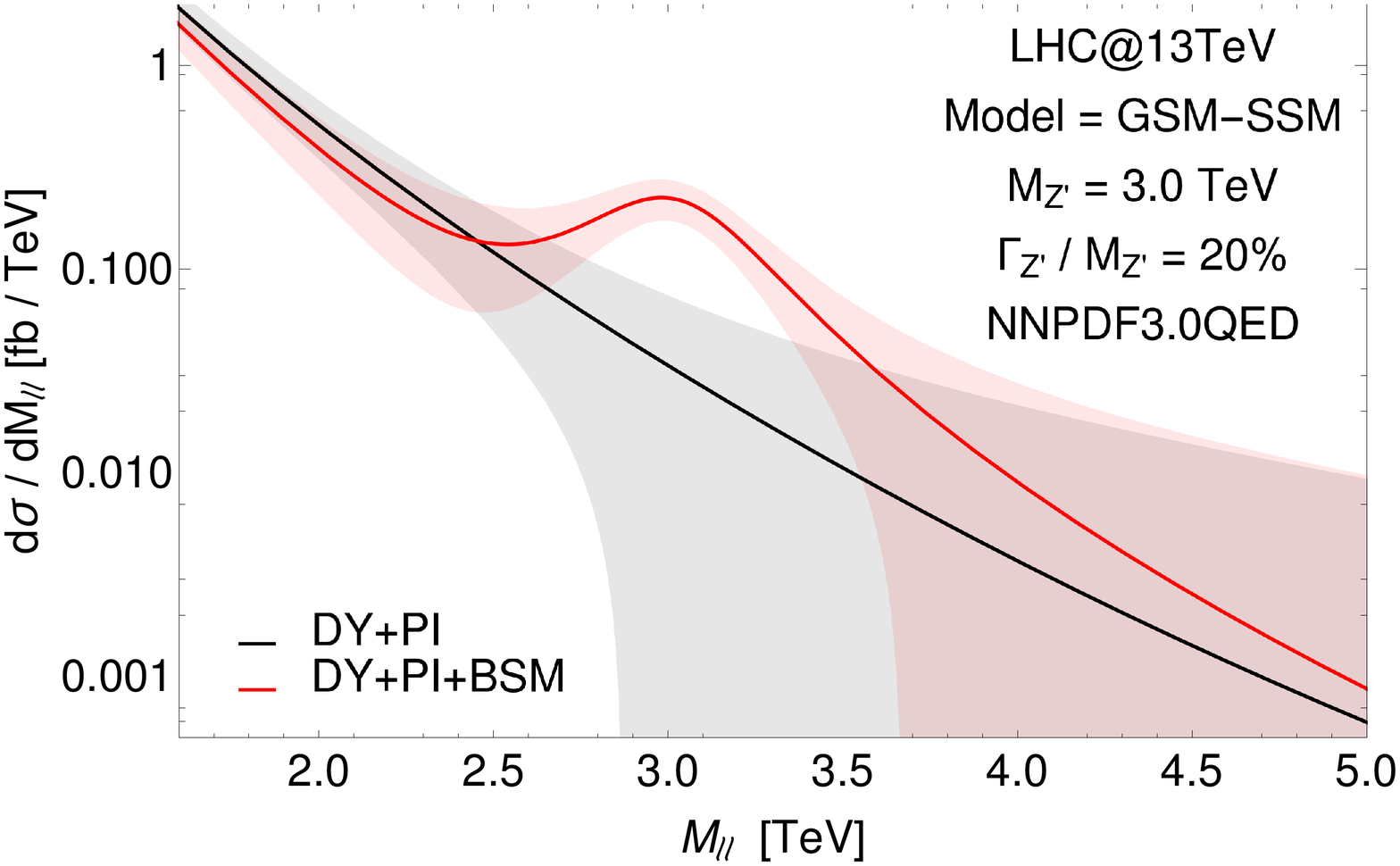}{(a)}
\includegraphics[width=0.45\textwidth]{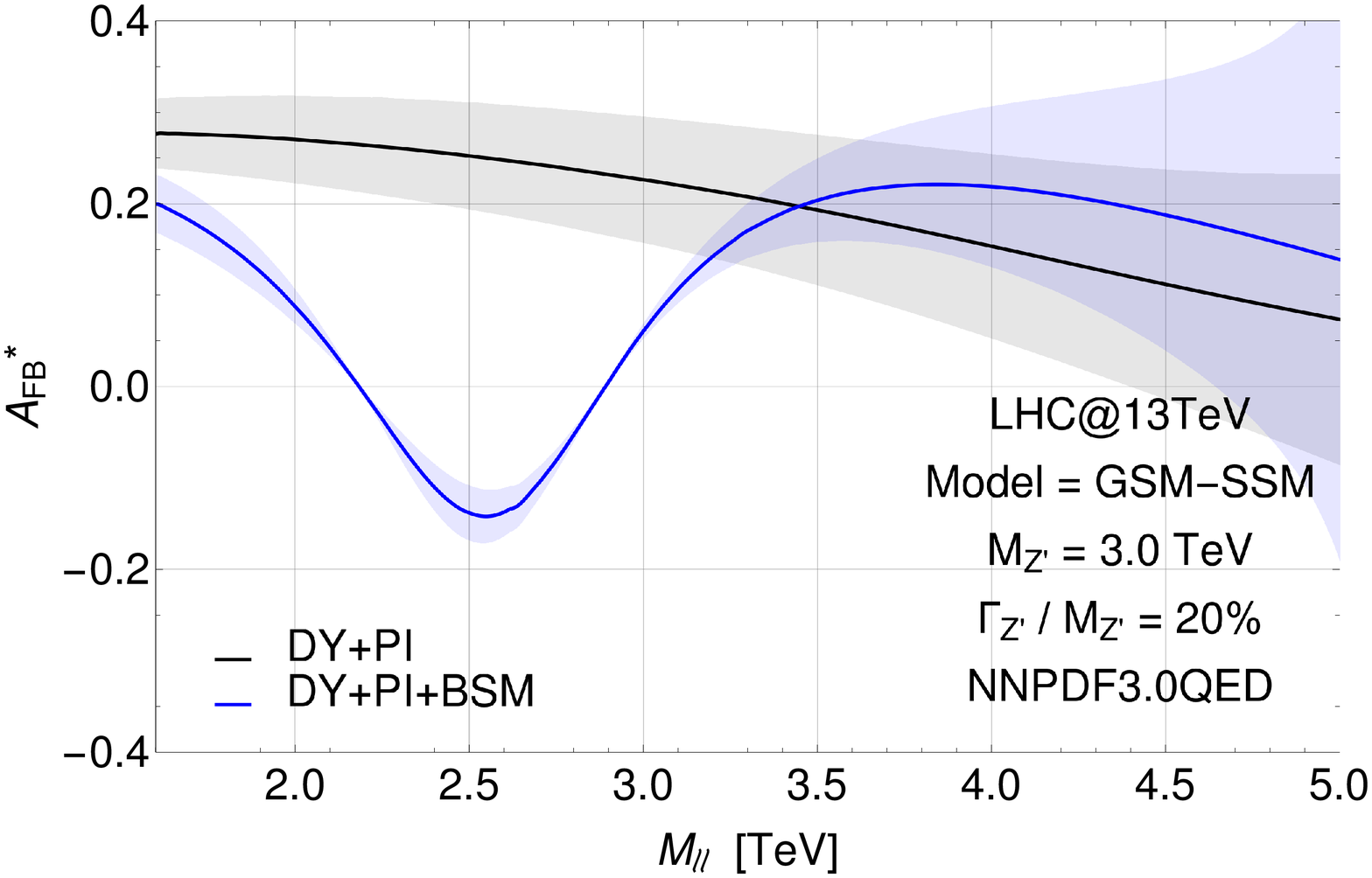}{(b)}
\includegraphics[width=0.45\textwidth]{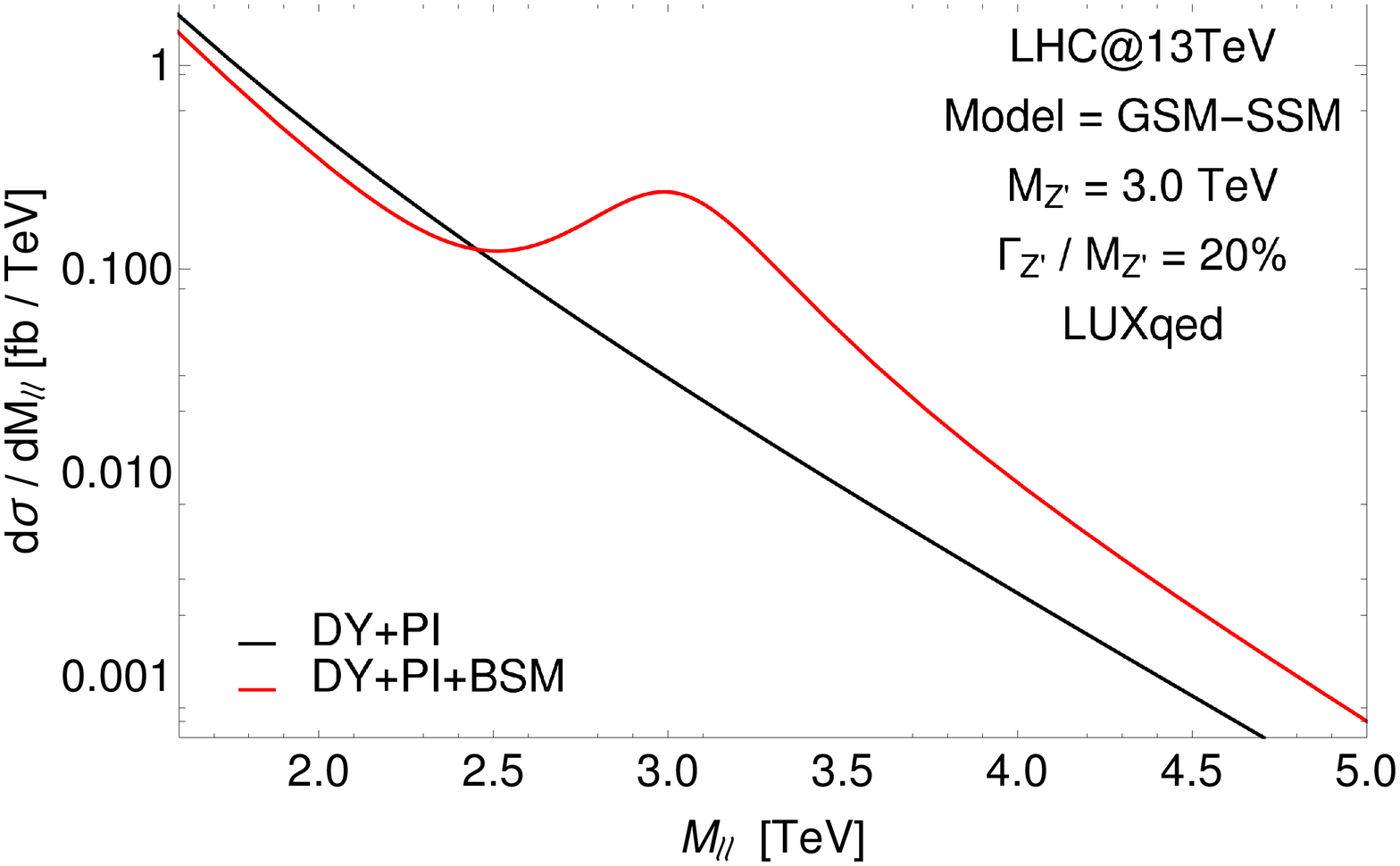}{(c)}
\includegraphics[width=0.45\textwidth]{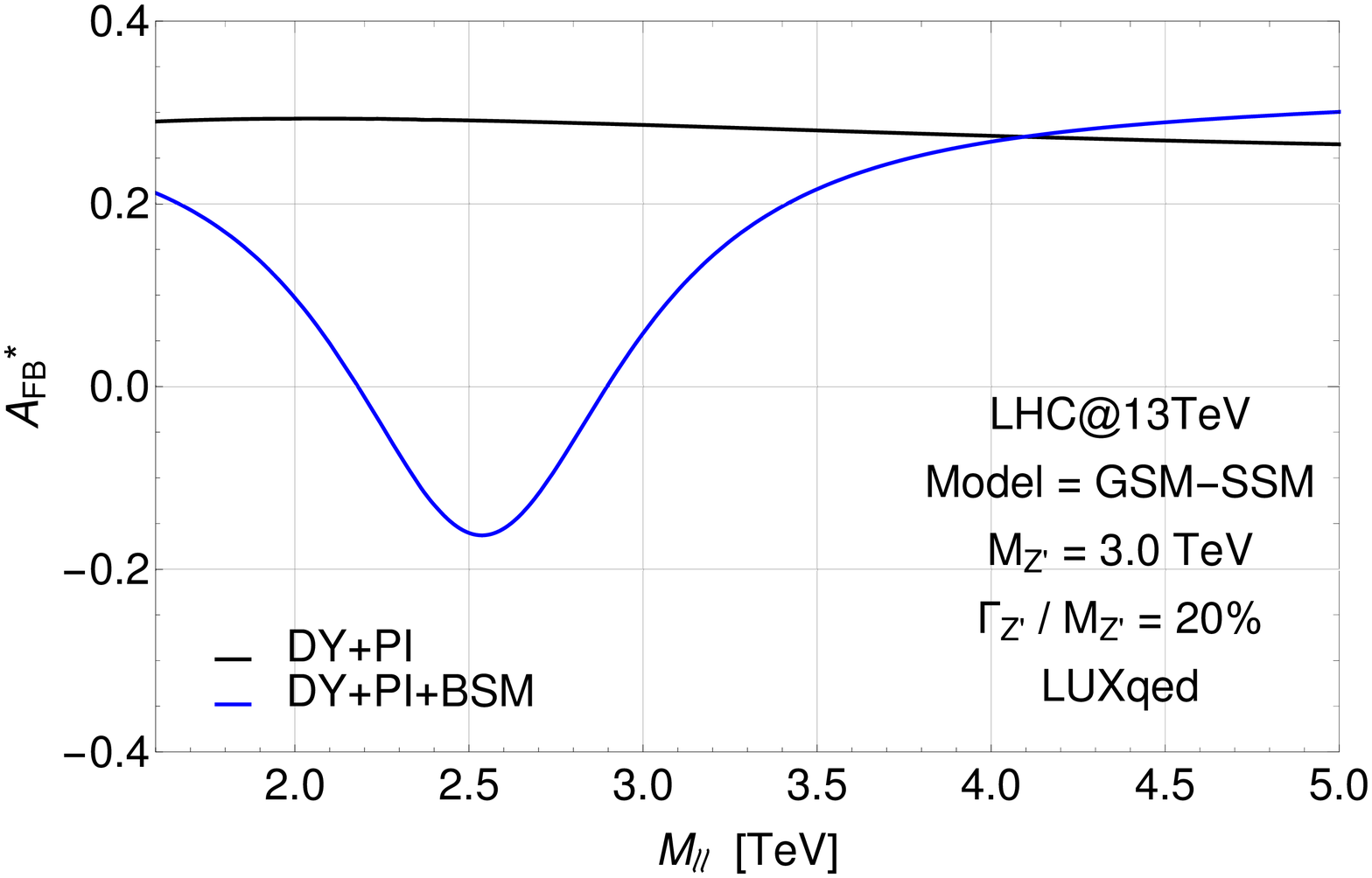}{(d)}
\caption{(a) Di-lepton spectrum and (b) Reconstructed $A_{FB}^*$ distribution at the LHC@13TeV in the $SSM$ with $M_{Z^\prime} = 3.0~TeV$ and $\Gamma_{Z^\prime} / M_{Z^\prime} = 20\%$, using NNPDF. 
The shaded areas represents the photon PDF error.
Standard acceptance cuts are applied: $|\eta_l| < 2.5$ and $p_T^l > 20~GeV$.
(c) and (d) same as (a) and (b) but with the use of the CTEQ6 set for the DY and the LUXqed set for the PI contributions. The photon PDF uncertainties are here too small to be visible.}
\label{fig:Zprime_signal}
\end{center}
\end{figure}

In Fig.~\ref{fig:Zprime_signal}, we show the di-lepton spectrum 
for the $SSM$ with
$M_{Z^\prime} = 3$ TeV and $\Gamma_{Z^\prime} / M_{Z^\prime} =
20\%$, considering both the cross section and reconstructed Forward-Backward Asymmentry ($A^*_{FB}$) distributions observables,
adopting the NNPDF3.0QED set (Fig.~\ref{fig:Zprime_signal}(a) and (b)) and the LUXqed (Fig.~\ref{fig:Zprime_signal}(c) and (d)). The latter is representative of the CT14QED\_inc result as well.
In computing the $Z^\prime$-boson signature, we take into account the
interference between the new heavy $Z^\prime$-boson and the SM $Z$ and
$\gamma$ as discussed in Refs.~\cite{Accomando:2013sfa,Accomando:2016sge,Accomando:2016mvz}. 
As visible in the two upper plots, the effects of the PI contribution and of its error are relevant in the NNPDF3.0QED case. 
PI events are decreasing our sensitivity to a non-resonant signal. In this specific case, computing the significance of the signal before and 
after the PI effects and systematics inclusion we would see the significance moving roughly from 3.7 to 2.8. 
On the other hand in the $A^*_{FB}$ observable the effect of the systematic uncertainty is reduced as expected (see Refs.~\cite{Accomando:2015cfa, Accomando:2015ava} and Ref.~\cite{Accomando:2016tah} for the case with the inclusion of photon PDFs).
\noindent
If we follow the LUXqed (or CT14QED\_inc) result (Fig.~\ref{fig:Zprime_signal}(c) and (d)) clearly the inclusion of PI processes do not have any effect,
thus in this scenario the PI component and its uncertainty can be safely neglected in di-lepton channel analysis.
\noindent
It is worth noting that the high PI central value predicted by the NNPDF collaboration has measurable effects in the $A^*_{FB}$ observable, especially in the high invariant mass region.
In general, being angularly symmetric, the inclusion of PI processes leads to a reduction of the overall SM predictions for this observable.
Comparing Fig.~\ref{fig:Zprime_signal}(b) and (d), we can see that in the NNPDF scenario the complete result deviates from the pure DY term, 
while it is  well represented by it in 
the LUXqed scenario in Fig.~\ref{fig:Zprime_signal}(d), as here the PI central value is negligible 
with respect to the DY.

\section{Conclusions}

Di-lepton final states in the high invariant mass region 
$ M_{l l } \geq 1 $ TeV are one of the primary channels for 
searches for $Z^\prime$ gauge bosons in BSM scenarios and 
for precision studies of the SM at the LHC. It was pointed out in 
\cite{Accomando:2016tah, Accomando:2016ouw} 
that in the high-mass region the contribution of photon-induced 
di-lepton production, and associated photon PDF uncertainties, 
could distort the di-lepton SM spectrum shape potentially affecting wide-resonance searches. 
This could have an even greater impact on the Contact Interaction (CI) type of search where one has to predict the SM background form theory. 
While for a standard narrow $Z^\prime$ search this is not a source of uncertainty, 
it can become one of the main sources of theoretical systematics for wide $Z^\prime$-bosons and CI searches.

In this letter we have extended the analysis of works \cite{Accomando:2016tah, Accomando:2016ouw} by, on one hand, 
including the contribution of virtual photon processes and, on the 
other hand, evaluating the impact of recent updates in photon PDF fits. 
\noindent
Using as benchmarks the SM di-lepton spectrum and the GSM-SSM 
wide-resonance scenario, we have found non-negligible contributions 
from the single-dissociative process. 
This points to the relevance of improving in the future the theory of 
PI processes at the LHC for both QED PDFs and elastic processes. 

We have presented a new study of the significance of the BSM signal 
in di-lepton channels by incorporating both real and virtual photon contributions. 
Our study examines  the need for including the theoretical systematics 
from PI production in the SM background estimate, depending 
on different scenarios for 
photon PDFs, and  illustrates 
the role  of 
analysing the reconstructed $A^*_{FB}$ in addition to the differential cross section.

\label{sec:summa}

\section*{Acknowledgements}
\noindent
We thank A. Belyaev and A. Pukhov for discussions on the EPA. We are grateful to V. Bertone for discussions on QED PDFs and for his help with NNPDF parton distribution calculations.
This work is supported by the Science and Technology Facilities Council, grant number ST/L000296/1. 
F. H. acknowledges the support and hospitality of DESY, the University of Hamburg and 
the DFG Collaborative Research Centre SFB 676 ``Particles, Strings and the Early Universe".  
All authors acknowledge partial financial support through the NExT Institute.

\bibliographystyle{apsrev4-1}
\bibliography{bib}

\end{document}